\documentclass[9pt,twocolumn,twoside]{osajnl}

\journal{ol} 

\setboolean{shortarticle}{true}

\title{Photonic spin Hall effect mediated by bianisotropy}

\author[1]{Dmitry~V.~Zhirihin}
\author[1,2]{Sergey~V.~Li}
\author[1]{Denis~Y.~Sokolov}
\author[1]{Alexey~P.~Slobozhanyuk}
\author[1*]{Maxim~A.~Gorlach}
\author[1,3,4] {Alexander~B.~Khanikaev}

\affil[1]{ITMO University, Saint Petersburg 197101, Russia}
\affil[2]{State Key Laboratory of Modern Optical Instrumentation, College of Optical Science and Engineering, Zhejiang University, Hangzhou 310027, China}
\affil[3]{The City College of the City University of New York, New York 10031, USA}
\affil[4]{Graduate Center of the City University of New York, New York, 10016, USA}

\affil[*]{Corresponding author: m.gorlach@metalab.ifmo.ru, khanikaev@gmail.com}

\begin{abstract}
Coupling of electric and magnetic responses of a scatterer known as bianisotropy enables rich physics and unique optical phenomena, including asymmetric absorption or reflection, one-way transparency, and photonic topological phases. Here we demonstrate yet another feature stemming from bianisotropic response, namely, polarization-dependent scattering of light by bianisotropic dielectric meta-atom with broken mirror symmetry, which yields a photonic analogue of spin Hall effect. Based on a simple dipole model, we explain the origin of the effect confirming our conclusions by experimental observation of photonic spin Hall effect both for a single meta-atom and for an array of them.
\end{abstract}

\setboolean{displaycopyright}{true}

\begin{document}

\newcommand{\red}[1]{{\color{red}#1}}
\newcommand{\eps}{\varepsilon}

\maketitle

Electric and magnetic resonances supported by dielectric particles offer unique functionalities in steering of the scattered light~\cite{Kuznetsov,Evlyukhin}. A particle with overlapping electric and magnetic dipole resonances excited by the incident wave may exhibit electric and magnetic dipole moments of comparable magnitude. Hence, the interference of the fields produces by these dipole moments can lead to the suppression of forward or backward scattering widely known as Kerker effect~\cite{Kerker}. Harnessing higher-order multipole resonances provides even greater flexibility in shaping scattering diagrams of nanoparticles~\cite{Alaee-OL,Liu}. Furthermore, by arranging Kerker-type scatterers into an array with subwavelength periodicity, one can design so-called Huygens' metasurface which, despite exhibiting a resonant excitation, transmits forward all incident radiation~\cite{Glybovski}, the phenomenon that has been recently verified experimentally at near infrared frequencies~\cite{Decker}.

An additional degree of freedom in light manipulation is associated with bianisotropic response of a scatterer. Bianisotropy is manifested through the coupling of the electric field to the magnetic dipole moment of the particle, and, by reciprocity, through the coupling of magnetic field to electric dipole moment of the particle~\cite{Semchenko}. Such cross-coupling is facilitated by the symmetry reduction of the scatterer, namely, breaking of the mirror symmetry. While the studies of bianisotropic structures and composites have quite a long history~\cite{Semchenko}, recent studies of bianisotropic metamaterials and metasurfaces~\cite{Asadchy-review,Odit} suggest a rich variety of new effects and applications ranging from peculiar features in scattering~~\cite{Alaee-Albooyeh,Alaee-Rockstuhl} to topological states in bianisotropic metamaterials~\cite{Khanikaev,Jacobs15,SlobNP}. In the former case, a particle with broken mirror symmetry illuminated along the symmetry axis exhibits completely different sets of excited multipoles for two opposite illumination directions~\cite{Alaee-Albooyeh,Alaee-Rockstuhl}, which causes different backscattering and different reflection phases, while in the latter case bianisotropy gives rise to the confinement of radiation to bianisotropic domain walls and robust guiding of light~\cite{Cheng-Khanikaev,Kang-Khanikaev,Purtseladze}.

\begin{figure}[h!]
\centering
\includegraphics[width=\linewidth]{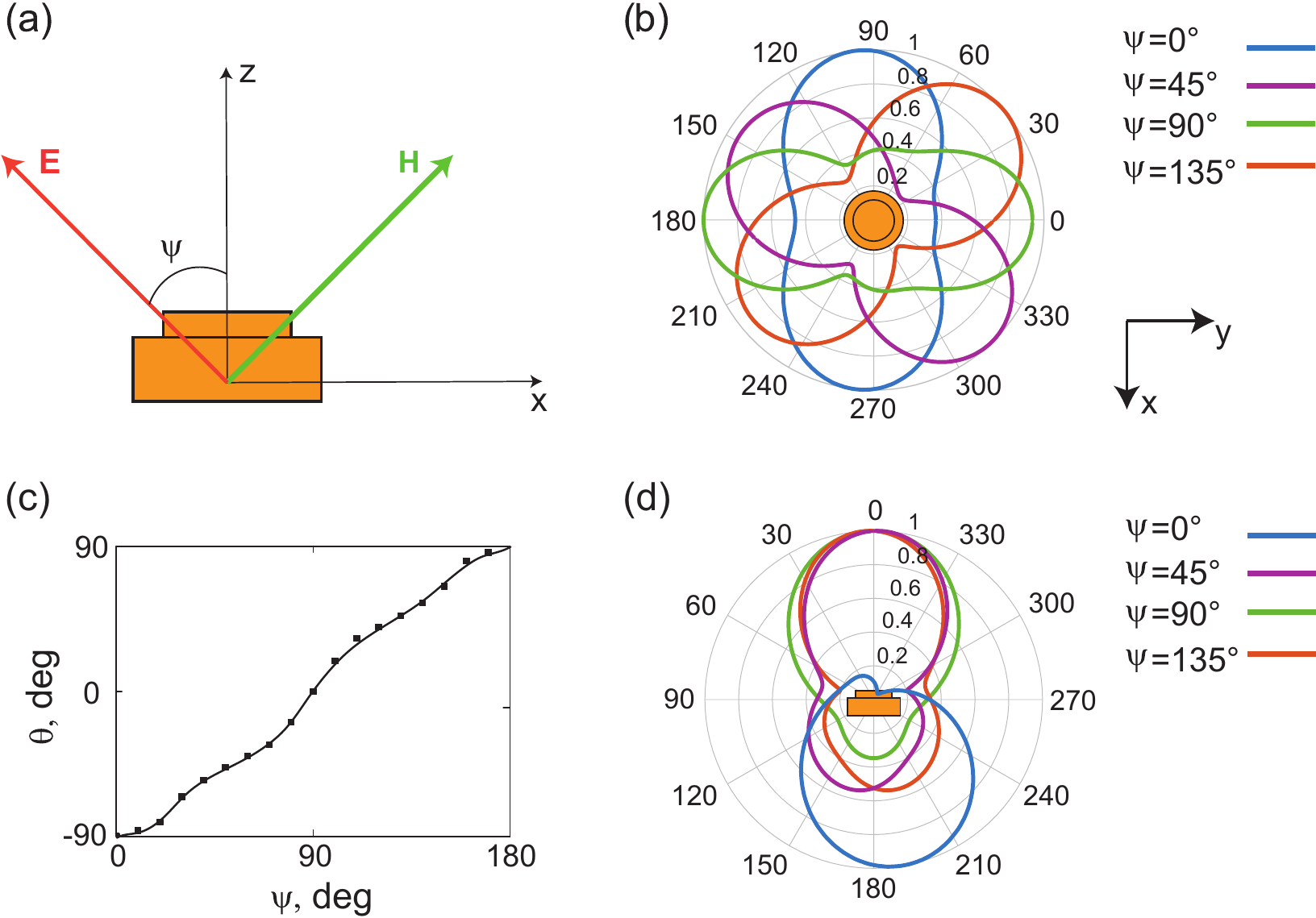}
\caption{Polarization dependence of light scattering. (a) Geometry of the scattering problem. $z$ axis coincides with the symmetry axis of the particle, $y$ axis is parallel to wave vector of the incident wave. $\psi$ is the angle between electric field of the incident linearly polarized wave and symmetry axis of the disk. (b) Scattering diagram of a bianisotropic meta-atom in $Oxy$ plane simulated in CST Microwave Studio software package for different linear polarizations of the incident wave. Legend shows the magnitude of polarization angle $\psi$ defined in panel (a). Numbers on circumference indicate the scattering angle $\varphi$. The magnitude of $|E|^2$ is plotted. Bianisotropic meta-atom consists of two ceramic disks with permittivity $\varepsilon=39$, diameters $d_1=29.1$~mm, $d_2=22.0$~mm and heights $h_1=9.0$~mm and $h_2=3.0$~mm attached to each other. (c) Dependence of the direction to the main lobe of scattering diagram in $Oxy$ plane on polarization of the incident wave. (d) Scattering diagram of a bianisotropic particle in the plane containing $z$ axis and orthogonal to the direction of the main scattering lobe in $Oxy$ plane for different polarizations of impinging wave.}
\label{fig:SpinHall}
\end{figure}

In this Letter, we make the next conceptual step and demonstrate polarization-dependent scattering of the electromagnetic wave on a single meta-atom with broken mirror symmetry. The particle is illuminated from the direction perpendicular to its symmetry axis in the geometry shown in Fig.~\ref{fig:SpinHall}(a). To describe the origin of polarization-dependent scattering, we start by developing a simplified dipole model assuming that the disk has electric and magnetic dipole responses only, thus neglecting higher order multipoles. With these assumptions, the response of the disk to the external electromagnetic field ${\bf E}$, ${\bf H}$ is described by the equations
\begin{gather}
{\bf d}=\hat{\alpha}^{\rm{ee}}\,{\bf E}+\hat{\alpha}^{\rm{em}}\,{\bf H}\:,\label{Dresponse}\\
{\bf m}=\hat{\alpha}^{\rm{me}}\,{\bf E}+\hat{\alpha}^{\rm{mm}}\,{\bf H}\:,\label{Mresponse}
\end{gather}
where ${\bf d}$ and ${\bf m}$ are electric and magnetic dipole moments of the disk, matrices $\hat{\alpha}$ denote $3\times 3$ polarizability tensors, and CGS system of units is employed. The structure of polarizability tensors is defined in turn by the symmetries of the meta-atom, namely, rotational axis $z$ and mirror plane $Oxz$. Specifically,
\begin{gather}
\hat{\alpha}^{\rm{ee}}=
\begin{pmatrix}
\alpha^{\rm{ee}} & 0 & 0 \\
0 & \alpha^{\rm{ee}} & 0 \\
0 & 0 & \alpha_{zz}^{\rm{ee}}
\end{pmatrix}
\:,\mspace{8mu}
\hat{\alpha}^{\rm{mm}}=
\begin{pmatrix}
\alpha^{\rm{mm}} & 0 & 0 \\
0 & \alpha^{\rm{mm}} & 0 \\
0 & 0 & \alpha_{zz}^{\rm{mm}}
\end{pmatrix}
\:,\mspace{8mu}\notag\\
\hat{\alpha}^{\rm{em}}=\hat{\alpha}^{\rm{me}}=
\begin{pmatrix}
0 & i\beta & 0\\
-i\beta & 0 & 0\\
0 & 0 & 0
\end{pmatrix}
\:.\label{Polarizabilities}
\end{gather}

\begin{figure}[ht]
\centering
\includegraphics[width=1.0\linewidth]{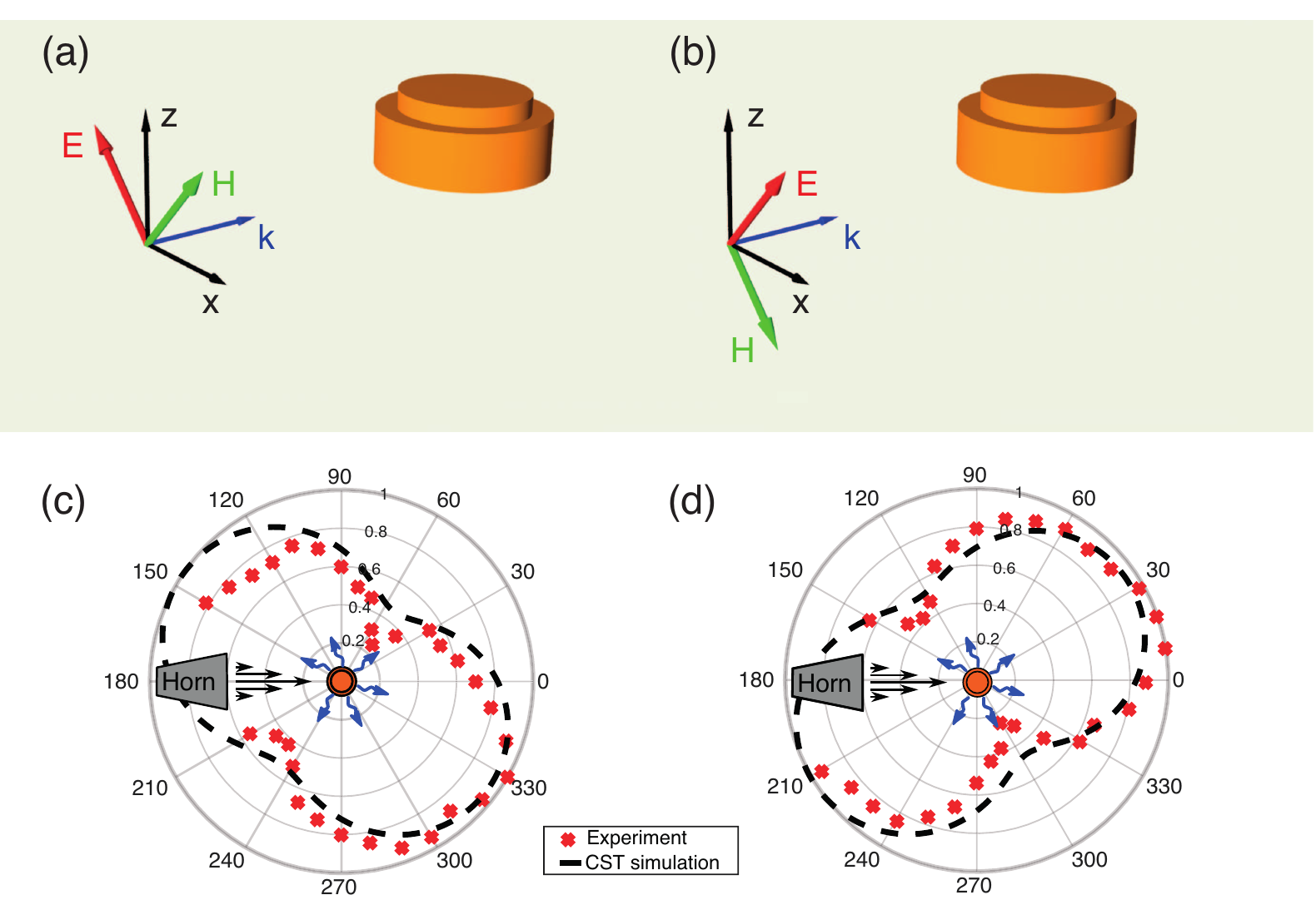}
\caption{Illustration of light scattering by a single meta-atom. (a,b) Geometry of the problem: electric field forms $\pm 45^\circ$ angle with the disk axis. (c,d) Scattering diagrams for geometries (a) and (b), respectively, showing the magnitude of $|E|$ for the scattered field calculated in $Oxy$ plane with CST Microwave Studio software. The frequency of impinging wave is set to $f=2.5$~GHz. Numbers on the circumference indicate the magnitude of $\varphi$ angle. Crosses indicate experimental results for the scattering diagram.}
\label{fig:SingleDisk}
\end{figure}

For simplicity, we assume that the disk can be polarized in $Oxy$ plane only, i.e. $\alpha_{zz}^{\rm{ee}}=\alpha_{zz}^{\rm{mm}}=0$. Then the impinging wave with polarization shown in Fig.~\ref{fig:SpinHall}(a) induces electric and magnetic dipole moments
\begin{gather}
{\bf d}=\left(-\alpha^{\rm{ee}}\,E_0\,\sin\psi,  -i\beta\,H_0\,\cos\psi, 0\right)^T\:,\\
{\bf m}=\left(\alpha^{\rm{mm}}\,H_0\,\cos\psi, i\beta\,E_0\,\sin\psi, 0\right)^T\:.
\end{gather}
where $E_0=H_0$ are the amplitudes of electric and magnetic field. Far field scattered by the disk in the direction specified by unit vector ${\bf n}=\left(-\sin\varphi, \cos\varphi, 0\right)^T$ at the distance $r$ from the disk center is given by~\cite{Landau2}
\begin{gather}
{\bf E}_{\rm{sc}}=\frac{q^2\,e^{iqr}}{r}\,\left[{\bf d}-{\bf n}\,({\bf n}\cdot{\bf d})\right]-\frac{q^2\,e^{iqr}}{r}\,\left[{\bf n}\times{\bf m}\right]\:,
\end{gather}
where $q=2\pi\,f/c$, $f$ being the frequency of the incident wave, $c$ being the speed of light. The intensity scattered by the disk in the direction specified by the unit vector ${\bf n}$ is quantified by $|{\bf E}_{\rm{sc}}|^2$
\begin{gather}\label{ScatteringDiagram}
|{\bf E}_{\rm{sc}}|^2\propto |{\bf d}|^2-|({\bf d}\cdot{\bf n})|^2+|{\bf m}|^2-|({\bf m}\cdot{\bf n})|^2\notag\:.
\end{gather}
or explicitly
\begin{gather}
|{\bf E}_{\rm{sc}}|^2\propto \left[|\alpha^{\rm{ee}}|^2\,\sin^2\psi+|\alpha^{\rm{mm}}|^2\,\cos^2\psi\right]\,\cos^2\varphi+|\beta|^2\,\sin^2\varphi-\notag\\
+\frac{1}{2}\,\text{Im}\,\left[\beta^*\,(\alpha^{\rm{ee}}+\alpha^{\rm{mm}})\right]\,\sin 2\psi\,\sin 2\varphi\:.\label{ScatteringFormula}
\end{gather}
At this point, we stress that even in the case of lossless disk the quantities $\alpha^{\rm{ee}}$, $\alpha^{\rm{mm}}$ and $\beta$ are complex because of the radiation loss correction to polarizabilities~\cite{Sipe,BelovPJTF}. Hence, the last term in \eqref{ScatteringFormula} is nonzero and captures the effect of scattering asymmetry with respect to the wave vector of the incident wave.  Furthermore, \eqref{ScatteringFormula} suggests that the scattering into $\varphi$ direction for the wave with $\psi$ angle of polarization is essentially the same as scattering into $2\pi-\varphi$ direction for the wave with $-\psi$ angle of polarization. In the other words, if the wave with polarization angle $\psi$ scatters predominantly to the right with respect to ${\bf k}$ direction, the wave with complementary polarization $-\psi$ will scatter predominantly to the left. In fact, such polarization-dependent scattering of light provides a photonic analogue of well-celebrated spin Hall effect~\cite{Yin-Zhang,Kapitanova,Lobanov}. Note, however, that in contrast to the previous studies, our proposal does not rely on gradient-index structures~\cite{Capasso} but rather employs the intrinsic asymmetry of scattering originating from bianisotropic response.

To verify the predictions of our simplified model, we have performed full-wave numerical simulations and calculated scattering diagrams for the bianisotropic dielectric meta-atom for different linear polarizations (i.e. different $\psi$) of the impinging wave. The results presented in Fig.~\ref{fig:SpinHall}(b) demonstrate that the scattering diagram rotates as the polarization of the incident wave is varied, such that the direction to the main lobe of the diagram changes gradually from $-90^\circ$ to $90^\circ$ as polarization angle is varied from $0^\circ$ to $180^\circ$ [Fig.~\ref{fig:SpinHall}(c)]. While the meta-atom exhibits a pronounced asymmetry of scattering in $Oxy$ plane, scattering in the perpendicular plane containing $z$ axis appears to be more symmetric [Fig.~\ref{fig:SpinHall}(d)] with a bit of forward-backward asymmetry stemming from interference of the fields from electric and magnetic dipoles.

\begin{figure}[h!]
\centering
\includegraphics[width=\linewidth]{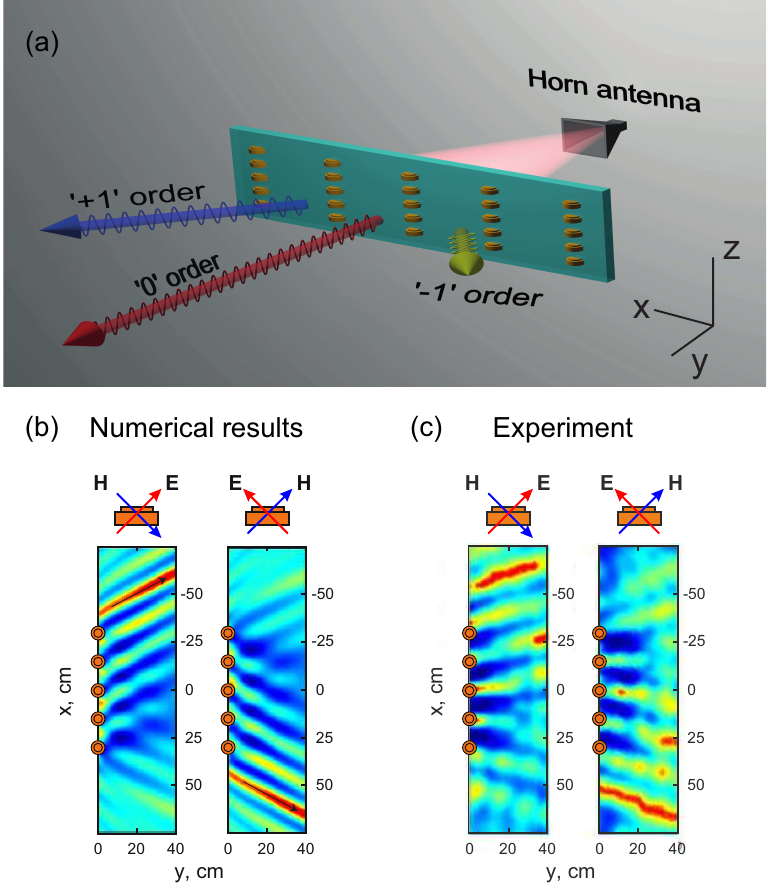}
\caption{(a) An artist view of the fabricated diffraction grating based on bianisotropic meta-atoms. Symmetry axis of the disk is parallel to the plane of  grating. Bianisotropy of the disks causes the asymmetry between +1 and -1 diffraction orders. (b) Simulated distribution of transmitted field intensity $|{\bf E}|^2$ for the diffraction grating made of bianisotropic meta-atoms at frequency $f=2.5$~GHz. Polarization angle of impinging wave is $\psi=45^\circ$ and $\psi=-45^\circ$ for left and right panels, respectively. Lattice periods along $x$ and $z$ axes are chosen to be $l_x=150$~mm and $l_z=30$~mm, respectively, such that the diffraction occurs only in $x$ direction and only $\pm 1$ diffraction orders appear. (c) Experimental colormap showing the intensity of transmitted field $|{\bf E}|^2$. Asymmetry of diffraction is clearly seen.}
\label{fig:Metasurface}
\end{figure}

A clear-cut demonstration of photonic spin Hall effect in our system is provided by the two representative cases $\psi=\pm 45^\circ$ [Fig.~\ref{fig:SingleDisk}(a,b)] when the asymmetry of scattering is so pronounced that the direction to the main scattering lobe has $\varphi_{\rm{lobe}}=\pm 39^\circ$  [Fig.~\ref{fig:SingleDisk}(c,d)]. To support our findings, we have fabricated bianisotropic meta-atoms made of high-permittivity ceramics ($\eps=39$)  and composed of two coaxial ceramic disks with diameters $d_1=29.1$~mm, $d_2=22.0$~mm and heights $h_1=9.0$~mm and $h_2=3.0$~mm closely attached to each other. The scatterer was placed on Styrofoam substrate and mounted on an azimuthal rotation setup. Measurements of scattering diagram were performed in an anechoic chamber in the range of angles from $30^\circ$ to $330^\circ$ using a pair of transmission and receiving antennas connected to vector network analyzer Rohde \& Schwarz ZVB20. The results of measurements presented in Fig.~\ref{fig:SingleDisk}(c,d) confirm the presence of polarization-dependent scattering in the system showing good agreement with the results of numerical simulations.

However, the signal from the individual disk is typically rather weak. To provide stronger evidence of photonic spin Hall effect, we have fabricated a periodic structure based on rectangular lattice of such bianisotropic meta-atoms. Scattering asymmetry inherent to a single particle will then be reproduced at the level of collective response and can be manifested, for instance, in the deflection of normally incident Gaussian beam. Moreover, such intriguing performance can also be detected when the array is illuminated by the plane wave. To this end, we choose the lattice constant in $x$ direction $l_x=150$~mm to be larger than the wavelength of operation ($\lambda\approx 120$~mm), while the lattice constant in $z$ direction $l_z=30$~mm is chosen to be subwavelength. By the proper choice of $l_x$ we ensure that, besides the trivial zeroth diffraction maximum, such diffraction grating exhibits also $\pm 1$ diffraction orders. The scattering asymmetry is then manifested as a difference in the intensities of $+1$ and $-1$ diffraction maxima as illustrated in Fig.~\ref{fig:Metasurface}(a).

As we have checked numerically, the intensities of $+1$ and $-1$  diffraction orders differ up to 20 times at the frequency $2.5$~GHz, while the intensities of $+1$ and $0$ diffraction orders appear to be comparable. Note that further optimization of the diffraction grating may allow one to fully suppress $-1$ and $0$ diffraction orders, thus rerouting all the transmitted energy into $+1$ diffraction order~\cite{Fan}, which can be realized even at optical frequencies~\cite{Khaidarov}.

As a consequence of the diffraction asymmetry, the distribution of transmitted field under the normal illumination will be strongly asymmetric [Fig.~\ref{fig:Metasurface}(b)]. Asymmetry in transmission is accompanied by hot-spot on the corresponding side of the diffraction grating.  To verify our conclusions, we have performed a proof-of-concept experiment with $5\times 5$ array of bianisotropic meta-atoms inserted into the Styrofoam substrate which is transparent at microwave frequencies. The grating was placed in an anechoic chamber and illuminated by the signal from TMA ultra-wideband horn antenna, while the field map was measured in the area $1.5\times 0.4~m^2$ using a subwavelength electric dipole mounted on a precision 3-axis mechanical scanner. The horn antenna and the electric probe were connected to the ports of vector network analyzer Rohde \& Schwarz ZVB20. The retrieved distribution of the transmitted field intensity is provided in Fig.~\ref{fig:Metasurface}(c), confirming the collective manifestation of photonic spin Hall effect in the system.

To summarize, we have demonstrated polarization-dependent routing of linearly polarized electromagnetic waves stemming from the interplay of electric and magnetic meta-atom responses supplemented by magneto-electric coupling both for individual bianisotropic scatterers and for the arrays of them. We have proved that tuning the angle between polarization plane of light and particle symmetry axis makes it is possible to  rotate arbitrarily the scattering diagram of such bianisotropic meta-atom. We believe that the observed bianisotropy-induced photonic spin Hall effect paves a way towards new applications, including metamaterials with topologically nontrivial photonic bands with the coupling of the elements controlled by polarization of an impinging wave.

This work was supported by the Russian Science Foundation (Grant No.~18-72-00102). M.A.G. acknowledges partial support by the Foundation for the Advancement of Theoretical Physics and Mathematics ``Basis". A.B.K. acknowledges the National Science Foundation grants EFRI-1641069 and DMR-1809915. Research was carried out in part at the Center for Functional Nanomaterials, Brookhaven National Laboratory, which is supported by the U.S. Department of Energy, Office of Basic Energy Sciences, under Contract No. DE-SC0012704.

\bibliography{SpinHall}

\begin{thebibliography}{10}
\newcommand{\enquote}[1]{``#1''}

\bibitem{Kuznetsov}
A.~I. Kuznetsov, A.~E. Miroshnichenko, Y.~H. Fu, J.~B. Zhang, and
  B.~Luk'yanchuk, \enquote{Magnetic light,} {\protect\JournalTitle{Scientific
  Reports}} \textbf{2}, 492 (2012).

\bibitem{Evlyukhin}
A.~B. Evlyukhin, S.~M. Novikov, U.~Zywietz, R.~L. Eriksen, C.~Reinhardt, S.~I.
  Bozhevolnyi, and B.~N. Chichkov, \enquote{{Demonstration of Magnetic Dipole
  Resonances of Dielectric Nanospheres in the Visible Region},}
  {\protect\JournalTitle{Nano Letters}} \textbf{12}, 3749--3755 (2012).

\bibitem{Kerker}
M.~Kerker, D.~S. Wang, and C.~L. Giles, \enquote{Electromagnetic scattering by
  magnetic spheres,} {\protect\JournalTitle{Journal of Optical Society of
  America}} \textbf{73}, 765--767 (1983).

\bibitem{Alaee-OL}
R.~Alaee, R.~Filter, D.~Lehr, F.~Lederer, and C.~Rockstuhl, \enquote{{A
  generalized Kerker condition for highly directive nanoantennas},}
  {\protect\JournalTitle{Optics Letters}} \textbf{40}, 2645--2648 (2015).

\bibitem{Liu}
W.~Liu and Y.~S. Kivshar, \enquote{{Generalized Kerker effects in nanophotonics
  and meta-optics},} {\protect\JournalTitle{Optics Express}} \textbf{26},
  13085--13105 (2018).

\bibitem{Glybovski}
S.~B. Glybovksi, S.~A. Tretyakov, P.~A. Belov, Y.~S. Kivshar, and C.~R.
  Simovski, \enquote{{Metasurfaces: From microwaves to visible},}
  {\protect\JournalTitle{Physics Reports}} \textbf{634}, 1--72 (2016).

\bibitem{Decker}
M.~Decker, I.~Staude, M.~Falkner, J.~Dominguez, D.~N. Neshev, I.~Brener,
  T.~Pertsch, and Y.~S. Kivshar, \enquote{{High-Efficiency Dielectric Huygens'
  Surfaces},} {\protect\JournalTitle{Advanced Optical Materials}} \textbf{3},
  813--820 (2015).

\bibitem{Semchenko}
I.~Semchenko, A.~Serdyukov, A.~Sihvola, and S.~Tretyakov,
  \emph{{Electromagnetics of Bi-Anisotropic Materials: Theory and
  Applications}} (Gordon and Breach Science, Philadelphia, 2001).

\bibitem{Asadchy-review}
V.~S. Asadchy, A.~Diaz-Rubio, and S.~A. Tretyakov, \enquote{{Bianisotropic
  metasurfaces: physics and applications},}
  {\protect\JournalTitle{Nanophotonics}} \textbf{7}, 1069--1094 (2018).

\bibitem{Odit}
M.~Odit, P.~Kapitanova, P.~Belov, R.~Alaee, C.~Rockstuhl, and Y.~S. Kivshar,
  \enquote{Experimental realisation of all-dielectric bianisotropic
  metasurface,} {\protect\JournalTitle{Applied Physics Letters}} \textbf{108},
  221903 (2016).

\bibitem{Alaee-Albooyeh}
R.~Alaee, M.~Albooyeh, M.~Yazdi, N.~Komjani, C.~Simovski, F.~Lederer, and
  C.~Rockstuhl, \enquote{{Magnetoelectric coupling in nonidentical plasmonic
  nanoparticles: Theory and applications},} {\protect\JournalTitle{Phys. Rev.
  B}} \textbf{91}, 115119 (2015).

\bibitem{Alaee-Rockstuhl}
R.~Alaee, M.~Albooyeh, A.~Rahimzadegan, M.~S. Mirmoosa, Y.~S. Kivshar, and
  C.~Rockstuhl, \enquote{All-dielectric reciprocal bianisotropic
  nanoparticles,} {\protect\JournalTitle{Phys. Rev. B}} \textbf{92}, 245130
  (2015).

\bibitem{Khanikaev}
A.~B. Khanikaev, S.~H. Mousavi, W.-K. Tse, M.~Kargarian, A.~H. MacDonald, and
  G.~Shvets, \enquote{Photonic topological insulators,}
  {\protect\JournalTitle{Nature Materials}} \textbf{12}, 233 (2012).

\bibitem{Jacobs15}
D.~A. Jacobs, A.~E. Miroshnichenko, Y.~S. Kivshar, and A.~B. Khanikaev,
  \enquote{{Photonic topological Chern insulators based on Tellegen
  metacrystals},} {\protect\JournalTitle{New J. Phys.}} \textbf{17}, 125015
  (2015).

\bibitem{SlobNP}
A.~Slobozhanyuk, S.~H. Mousavi, X.~Ni, D.~Smirnova, Y.~S. Kivshar, and A.~B.
  Khanikaev, \enquote{Three-dimensional all-dielectric photonic topological
  insulator,} {\protect\JournalTitle{Nature Photonics}} \textbf{11}, 130--136
  (2017).

\bibitem{Cheng-Khanikaev}
X.~Cheng, C.~Jouvaud, X.~Ni, S.~H. Mousavi, A.~Z. Genack, and A.~B. Khanikaev,
  \enquote{Robust reconfigurable electromagnetic pathways within a photonic
  topological insulator,} {\protect\JournalTitle{Nature Materials}}
  \textbf{15}, 542--548 (2016).

\bibitem{Kang-Khanikaev}
Y.~Kang, X.~Ni, X.~Cheng, A.~B. Khanikaev, and A.~Z. Genack,
  \enquote{{Pseudo-spin-valley coupled edge states in a photonic topological
  insulator},} {\protect\JournalTitle{Nature Communications}} \textbf{9}, 3029
  (2018).

\bibitem{Purtseladze}
X.~Ni, D.~Purtseladze, D.~A. Smirnova, A.~Slobozhanyuk, A.~Al{\`u}, and A.~B.
  Khanikaev, \enquote{{Spin and valley-polarized one-way Klein tunneling in
  photonic topological insulators},} {\protect\JournalTitle{Science Advances}}
  \textbf{4}, eaap8802 (2018).

\bibitem{Landau2}
L.~D. Landau and E.~M. Lifshitz, \emph{{Course of Theoretical Physics. Vol.~2.
  The Classical Theory of Fields}} (Butterworth-Heinemann, Heidelberg, 1994),
  4th ed.

\bibitem{Sipe}
J.~E. Sipe and J.~V. Kranendonk, \enquote{Macroscopic electromagnetic theory of
  resonant dielectrics,} {\protect\JournalTitle{Physical Review A}} \textbf{9},
  1806--1822 (1974).

\bibitem{BelovPJTF}
P.~A. Belov, S.~I. Maslovski, K.~R. Simovski, and S.~A. Tretyakov, \enquote{{A
  Condition Imposed on the Electromagnetic Polarizability of a Bianisotropic
  Lossless Scatterer},} {\protect\JournalTitle{Technical Physics Letters}}
  \textbf{29}, 718--720 (2003).

\bibitem{Yin-Zhang}
X.~Yin, Z.~Ye, J.~Rho, Y.~Wang, and X.~Zhang, \enquote{{Photonic spin Hall
  effect at metasurfaces},} {\protect\JournalTitle{Science}} \textbf{339},
  1405--1407 (2013).

\bibitem{Kapitanova}
P.~V. Kapitanova, P.~Ginzburg, F.~J. Rodriguez-Fortu{\~n}o, D.~S. Filonov,
  P.~M. Voroshilov, P.~A. Belov, A.~Poddubny, Y.~Kivshar, G.~A. Wurtz, and
  A.~V. Zayats, \enquote{{Photonic spin Hall effect in hyperbolic metamaterials
  for polarization-controlled routing of subwavelength modes},}
  {\protect\JournalTitle{Nature Communications}} \textbf{5}, 3226 (2014).

\bibitem{Lobanov}
S.~V. Lobanov, S.~G. Tikhodeev, N.~A. Gippius, A.~A. Maksimov, E.~V. Filatov,
  I.~I. Tartakovskii, V.~D. Kulakovskii, W.~T., S.~C., J.~Gebler, M.~Kamp, and
  S.~H{\"o}fling, \enquote{Controlling circular polarization of light emitted
  by quantum dots using chiral photonic crystal slabs,}
  {\protect\JournalTitle{Phys. Rev. B}} \textbf{92}, 205309 (2015).

\bibitem{Capasso}
N.~Yu, P.~Genevet, M.~A. Kats, F.~Aieta, J.-P. Tetienne, F.~Capasso, and
  Z.~Gaburro, \enquote{{Light Propagation with Phase Discontinuities:
  Generalized Laws of Reflection and Refraction},}
  {\protect\JournalTitle{Science}} p. 1210713 (2011).

\bibitem{Fan}
Z.~Fan, M.~R. Shcherbakov, M.~Allen, J.~Allen, and G.~Shvets, \enquote{{Perfect
  Diffraction with Bianisotropic Metagratings},} {\protect\JournalTitle{arXiv:
  1802.01269}}  (2018).

\bibitem{Khaidarov}
E.~Khaidarov, H.~Hao, R.~Paniagua-Dominguez, Y.~F. Yu, Y.~H. Fu, V.~Valuckas,
  S.~L.~K. Yap, Y.~T. Toh, J.~S.~K. Ng, and A.~I. Kuznetsov,
  \enquote{{Asymmetric Nanoantennas for Ultrahigh Angle Broadband Visible Light
  Bending},} {\protect\JournalTitle{Nano Letters}} \textbf{17}, 6267--6272
  (2017).

\end{thebibliography}

\bibliographyfullrefs{SpinHall}


%

\end{document}